\documentclass[aps,twocolumn]{revtex4-1}
\usepackage{epsfig}
\usepackage{graphicx}
\usepackage{amsmath,amssymb,amsfonts}
\usepackage{array}
\usepackage{url}
\usepackage{hyperref}
\usepackage{multirow}
\usepackage{float}
\usepackage{lineno}
\usepackage{xspace}
\usepackage{caption}
\usepackage{subcaption}
\usepackage{ulem}
\usepackage{csquotes}
\usepackage[usenames,dvipsnames]{color}

\newcommand{\bef}{\begin{figure}}
\newcommand{\eef}{\end{figure}}
\newcommand{\bc}{\begin{center}}
\newcommand{\ec}{\end{center}}

\newcommand{\be}{\begin{equation}}
\newcommand{\ee}{\end{equation}}
\newcommand{\bea}{\begin{eqnarray}}
\newcommand{\eea}{\end{eqnarray}}

\def\ba{\begin{eqnarray}}
\def\ea{\end{eqnarray}}
\begin{document}

\title{$J/\psi$ and $\psi$(2S) polarization in proton-proton collisions at energies available at the CERN Large Hadron Collider using PYTHIA8}

 
\author{Bhagyarathi Sahoo$^1$}
\email{Bhagyarathi.Sahoo@cern.ch}
\author{Dushmanta Sahu$^1$}
\email{Dushmanta.Sahu@cern.ch}
\author{Suman Deb$^2$}
\email{sumandeb0101@gmail.com}
\author{Captain R. Singh$^1$}
\email{captainriturajsingh@gmail.com}
\author{Raghunath Sahoo$^1$}
\email{Corresponding Author: Raghunath.Sahoo@cern.ch}
\affiliation{$^1$Department of Physics, Indian Institute of Technology Indore, Simrol, Indore 453552, India}

\affiliation{$^2$Laboratoire de Physique des 2 infinis Irène Joliot-Curie, Université Paris-Saclay, CNRS-IN2P3, F-91405 Orsay, France}

\begin{abstract}
The production mechanisms of charmonium states in both hadronic and heavy-ion collisions hold great significance for investigating the hot 
and dense QCD matter. Studying charmonium polarization in ultra-relativistic collisions can also provide 
insights into the underlying production mechanisms. With this motivation, we explore the  $J/\psi$ and $\psi$(2S) 
polarization in proton+proton collisions at $\sqrt{s}$ = 7, 8, and 13 TeV using a pQCD-inspired Monte-Carlo event generator called 
PYTHIA8. This work considers reconstructed quarkonia through their dimuons decay channel in the ALICE forward 
rapidity acceptance range of $2.5 < y_{\mu \mu} < 4$. Further, we calculate the polarization parameters $\lambda_{\theta}$, 
$\lambda_{\phi}$, $\lambda_{\theta \phi}$ from the polar and azimuthal angular distributions of the dimuons in helicity and Collins-Soper frames. This study presents a comprehensive measurement of the polarization parameters as a function of transverse momentum, charged-particle 
multiplicity, and rapidity at the LHC energies. Our findings of charmonium polarization are in qualitative agreement with the 
corresponding experimental data.

\end{abstract}
\date{\today}
\maketitle

\section{Introduction}
\label{intro}

Despite being discovered nearly five decades ago, heavy quarkonia states remain a challenging puzzle for QCD-based 
models~\cite{Brambilla:2010cs}, due to their non-relativistic nature and the complex multi-scale dynamics involved in ultra-relativistic heavy-ion collisions (HICs). Several theoretical models have been developed to comprehend the quarkonium production 
mechanisms; the non-relativistic quantum chromodynamics (NRQCD) is one such model~\cite{Bodwin:1994jh}. The color octet 
NRQCD~\cite{Kramer:2001hh} explains the quarkonium production cross section and matches with the experimental data of high energy 
collider experiments such as Tevatron~\cite{CDF:1997ykw, CDF:1999fhr, D0:1996awi}, RHIC~\cite{PHENIX:2019ihw, 
STAR:2019vkt,Trzeciak:2014cma}, and the LHC~\cite{ ALICE:2017leg, ALICE:2019pid, LHCb:2015foc, CMS:2017dju, ATLAS:2015zdw, ATLAS:2012lmu, 
LHCb:2018yzj, ALICE:2015pgg}. In comparison, the color singlet model (CSM) of NRQCD~\cite{Baier:1981uk, Berger:1980ni, Chang:1979nn, Einhorn:1975ua, Kartvelishvili:1978id} predicts  $J/\psi$ and $\psi$(2S) production 
cross-sections 50 times smaller than the experimental observation at CDF collaboration in proton+proton ($pp$) collisions at $\sqrt{s}$ = 1.8 
TeV~\cite{CDF:1997ykw}. There are other various factorization techniques to predict the production cross-section of quarkonium, 
such as NRQCD factorization~\cite{Bodwin:1994jh, Bodwin:2013nua}, leading power fragmentation, next-to-leading-power fragmentation, 
the color singlet model, the color evaporation 
model (CEM)~\cite{Gluck:1977zm, Fritzsch:1977ay, Halzen:1977rs}, and the $k_T$-factorization approach~\cite{Baranov:2016clx, 
Braaten:2014ata}. The NRQCD calculations use color-octet matrix elements to account for the non-perturbative long-distance physics 
in heavy quarkonium systems. These matrix elements are adjusted to explain experimental data, improving the 
agreement between theory and experiment. However, this adjustment introduces model dependence and should be interpreted cautiously. 
Including next-to-leading-order (NLO) QCD corrections in CSM, the quarkonium production rates have shown a significant 
increase in the large transverse momentum ($p_{\rm T}$) region, i.e., $p_{\rm T} >$ 20 GeV~\cite{Lansberg:2008gk}. This increase has led to a notable 
reduction in the required contributions from color octet components to match the measured quarkonium production cross-section at the Tevatron~\cite{Lansberg:2008gk}.\\

However, for a comprehensive understanding of the quarkonium production mechanism, it is crucial to explore the dynamics responsible for its polarization. Studying the quarkonium polarization provides valuable information, e.g., quarkonium production mechanisms in $pp$ collisions, the effect of the deconfined medium on the formation of a bound state of two heavy quarks, and the role of spin-vorticity coupling in a thermal rotating medium~\cite{Faccioli:2022peq, DeMoura:2023jzz}, etc. Polarization refers to the alignment of the quarkonium spin with respect to a chosen axis in a reference frame. The details about the chosen reference frames are discussed in Sec.~\ref{formulation}. \\

Quarkonium polarization is predominantly investigated through the dilepton decay channel in experimental studies. The polarization of 
quarkonium states is obtained by analyzing the angular distributions of decay products. So far, from the experimental side, quarkonium 
polarization is observed at Tevatron~\cite{CDF:2000pfk, CDF:2001fdy, D0:2008yos, CDF:2007msx, CDF:2011ag}, RHIC~\cite{PHENIX:2009ghc, 
STAR:2013iae, STAR:2020igu}, and LHC~\cite{ALICE:2011gej, ALICE:2018crw, ALICE:2020iev,ALICE:2023jad, Etzion:2009zz, LHCb:2013izl, LHCb:2014brf,
CMS:2013gbz,CMS:2016xpm, CMS:2012bpf} at collider experiments, as well as at fixed target experiments such as E866 
(NuSea)~\cite{NuSea:2003fkm, NuSea:2000vgl} and HERA-B~\cite{HERA-B:2009iab}. From a theoretical standpoint, the polarization study has 
been discussed in Ref.~\cite{Gong:2012ug, Chao:2012iv,Ma:2018qvc, Butenschoen:2012px, Beneke:1998re, Haberzettl:2007kj, Shao:2014yta, 
Faccioli:2010kd,Shao:2012fs, Faccioli:2010ji, Faccioli:2010ej, Faccioli:2008dx, Beneke:1996yw, Braaten:1999qk, Kniehl:2000nn,Leibovich:1996pa, Gong:2008sn}. The 
color-octet model of NRQCD successfully explains the quarkonium production cross section but 
 fails to account for the polarization results of $J/\psi$ obtained by the CDF experiment at $\sqrt{s}$ = 1.96 
 TeV~\cite{CDF:2007msx}. It predicts that at very high momenta, quarkonia are produced from the fragmentation of gluons, 
 preserving their natural spin alignment~\cite{Braaten:1993rw, Faccioli:2010kd}. Therefore a large transverse polarization of 
 $J/\psi$ is estimated with respect to their momentum direction~\cite{Faccioli:2014cqa}. In addition, the leading order (LO) calculation in CSM of NRQCD predicts a strong transverse polarization for $J/\psi$ ~\cite{Gong:2008sn}, which also fails to explain the polarization data of $J/\psi$ at the LHC, RHIC and Tevatron~\cite{Butenschoen:2012px, Chung:2009xr}. While the inclusion of NLO calculation to the CSM  predicts a strong longitudinal polarization~\cite{Gong:2008sn, LHCb:2013izl, LHCb:2014brf, Lansberg:2008gk}. The effective theory NRQCD includes all the important contributions from both color-octet and color singlet intermediate states and gives a detailed overview of $J/\psi$ polarization~\cite{Butenschoen:2012px, Gong:2012ug,Chao:2012iv}.
 Furthermore, the quarkonium polarization parameters are estimated in $pp$, $p-A$, and $A-A$ collisions using an improved CEM employing the $k_{T}$ factorization~\cite{Cheung:2018tvq, Cheung:2018upe} and the collinear factorization~\cite{ Cheung:2021epq, Cheung:2022nnq} approach. It predicts either zero or slightly transverse polarization at high-$p_{\rm{T}}$ and a small longitudinal polarization at low-$p_{\rm{T}}$ depending on the polarization frame~\cite{Cheung:2021epq, Cheung:2022nnq, Vogt:2019zmr}. 
 On the other hand, with the current statistics, the experimental data of ALICE show a little or zero polarization for $J/\psi$ within uncertainty in both hadronic and nucleus-nucleus collisions, although LHCb 
predicts a small longitudinal polarization in the helicity frame. This discrepancy between theory and experiment is commonly called the \enquote{J/$\psi$ polarization puzzle}~\cite{ALICE:2011gej, 
Ma:2018qvc}. However, a Color Glass Condensate (CGC)+NRQCD approach~\cite{Ma:2018qvc} provides a qualitative description of the experimental 
data towards high $p_{\rm T}$ ( $p_{\rm T} > 4 $ GeV)at the LHC and STAR in minimum bias $pp$ collisions at forward rapidities~\cite{STAR:2020igu}.\\

In this work, we attempt to understand the charmonia polarization in $pp$ collisions by studying the $\lambda$-polarization 
parameters; $\lambda_{\theta}$, $\lambda_{\phi}$, and $\lambda_{\theta \phi}$. These parameters are obtained using PYTHIA8 simulation by 
taking the angular distribution of dimuons produced from $J/\psi$ and $\psi$(2S). Further, these $\lambda$-polarization parameters are 
studied as functions of $p_{\rm{T}}$, charged-particle multiplicity ($N_{ch}$), and rapidity ($y_{\mu \mu}$) corresponding to $\sqrt{s}$ = 7, 
8, and 13 TeV collision energies. The experimental measurement of quarkonium decay angular distribution is challenging because it demands 
a large number of event samples and a high level of accuracy in the subtraction of various kinematic correlations induced by detector 
acceptance. Thus, it is difficult to analyze the charged particle multiplicity and rapidity dependence of polarization parameters 
from the angular distribution of decay muons. Such difficulties can be easily overcome in the Monte-Carlo (MC) simulation studies. With 
the present understanding of $pp$ collisions dynamics at the LHC energies, the charged-particle multiplicity dependence study of quarkonium 
polarization would be an interesting topic to investigate the medium effect, and it may serve as a benchmark for heavy-ion 
collisions as well. The $p_{\rm{T}}$-dependence of the polarization parameter may help us to understand the dynamics of the particle production.  
Similarly, the rapidity dependence of polarization studies may reveal the phase-space analysis of the particles produced in ultra-relativistic collisions.\\

This paper is organized as follows. The brief details of dimuon angular distribution and event generation are described in 
Sec.~\ref{formulation}. Section~\ref{result} includes the results obtained by analyzing the angular distribution of dimuons 
in $pp$ collisions at the LHC energies. Section~\ref{result} consists of three subsections. The transverse momentum dependence of 
polarization parameters is discussed in Sec.~\ref{pt}. In Sec.~\ref{mul}, the charged-particle multiplicity dependence 
of polarization parameters is studied. The rapidity dependence study of polarization parameters is discussed in 
Sec.~\ref{rap}. Finally, the important findings are summarized, and the future scopes are presented in Sec.~\ref{summary}.\\

\begin{widetext}
    
\end{widetext}
\begin{figure}[!]    

   \includegraphics[scale=0.5]{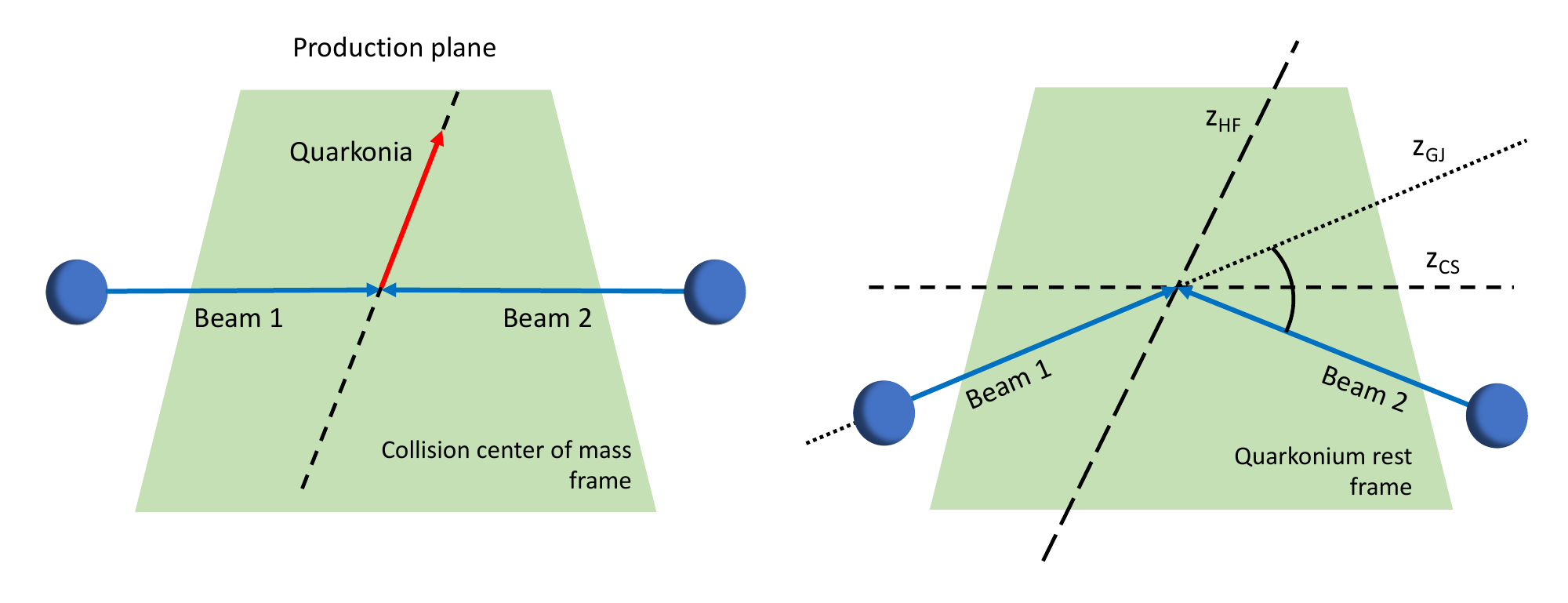}  
   
\caption{(Color online) Illustration of the three different definitions of the polarization axis, z, in the helicity (HF), Collins-Soper (CS), Gottfried-Jackson (GJ) reference frames, with respect to the direction of motion of the colliding beams (Beam 1 and Beam 2) and of the Quarkonia.}
 \label{fig:0}
\end{figure}

\section{ Analysis Methodology and Event generation}

\label{formulation}

\subsection{Dimuon decay angular distribution}
\label{dilepdist}
The polarization of the $J^{PC}= 1^{--}$ quarkonium states can be measured through the study of the angular distribution W($\theta, \phi$) of decay daughters in the dimuons decay channel and  can be parameterized as~\cite{Faccioli:2010kd,ALICE:2020iev,Braaten:2014ata};

\begin{widetext}
    \begin{equation}
        W(\theta,\phi) \propto \frac{1}{3+\lambda_{\theta}}  
    \left( 1+ \lambda_{\theta} \cos^{2}\theta + \lambda_{\phi} \sin^{2}\theta \cos2\phi + \lambda_{\theta \phi} \sin2\theta \cos\phi \right)
    \label{eq1}
    \end{equation}
\end{widetext}

where, $\theta$ and $\phi$ are the polar and azimuthal angles of the $\mu^{+}$  with respect to the spin-quantization axis (say, z-axis) of the chosen polarization frames and $\lambda_{\theta}$, $\lambda_{\phi}$, $\lambda_{\theta \phi}$ are the polarization parameters. In particular, the two cases ($\lambda_{\theta}$ = 1, $\lambda_{\phi}$ =0 , $\lambda_{\theta \phi}$ = 0 ) and ($\lambda_{\theta}$ = -1, $\lambda_{\phi}$ =0, $\lambda_{\theta \phi}$ = 0 ) correspond to the transverse and longitudinal polarization, respectively. The case ($\lambda_{\theta}$ = 0, $\lambda_{\phi}$ =0, $\lambda_{\theta \phi}$ = 0) correspond to zero polarization~\cite{ALICE:2020iev}.

There are three different conventions to define the polarization reference frames (definition of the z-axis), which are illustrated in Fig.~\ref{fig:0}~\cite{Faccioli:2010kd}.

\begin{itemize}
    \item  Helicity frame: In the direction of $J/\psi$ (or $\psi$(2S)) momentum in the center of the mass frame of the colliding beams.
    
    \item Collins-Soper frame: The bisector of the angle between the momentum of one beam and the successive direction of the other beam~\cite{Collins:1977iv}.
    
    \item Gottfried-Jackson frame: The direction of the momentum of one of the colliding beams~\cite{Gottfried:1964nx}.
\end{itemize}

The quantization axis of the Gottfried-Jackson frame lies in between the 
helicity and Collins-Soper reference frame~\cite{Faccioli:2010kd, 
Sarkar:2010zza}. So, in the current work, we solely cover the helicity 
and Collins-Soper frames as two extreme cases that are physically 
relevant. It is noteworthy to mention that the default setting for 
quarkonium production in hadronic collisions in all existing MC 
generators use an isotropic dilepton distribution, which is discussed in 
Ref.~\cite{Faccioli:2010kd}. In NA38, NA50, NA51, and NA60 experiments have 
measured a flat $\cos\theta_{\rm CS}$ ($\cos\theta$ distribution in 
Collins-Soper frame) angular distribution in the window $ |\rm 
\cos\theta_{\rm CS}| <$ 0.5 covering 50 \% of the phase space and assumed 
that  $J/\psi$ is unpolarized~\cite{ Sarkar:2010zza}. But, the recent 
global analysis of $J/\psi$ polarization measurement indicates that the 
$J/\psi$ is significantly polarized and its polarization changes 
longitudinal to transverse from low $p_{\rm{T}}$ to high 
$p_{\rm{T}}$~\cite{ALICE:2011gej}. We use PYTHIA8 to visualize the 
angular distribution of decay muons in Fig.~\ref{fig:1}, in order to validate the mentioned assumption. From Fig.~\ref{fig:1}, it is observed that the $\cos\theta$ angular distribution is almost isotropic in the angular range $ |\rm 
\cos\theta| <$ 0.5, however, the deviation from isotropic distribution starts at $|\rm \cos\theta| >$ 0.5 for both frames of reference. This is the consequence of physics processes involved in PYTHIA8, such as the production and decay of higher excited resonances, and the emission of gluons in the final state radiations, etc.~\cite{manual}. The distribution of dimuons with 
uniform acceptance and efficiency over $\cos\theta$ and $\phi$ 
distribution at the generation level allows us to determine the 
polarization parameters from the observed angular 
distribution~\cite{Faccioli:2010kd, Faccioli:2022peq}.

\begin{equation}
      <\cos^{2}\theta> = \frac{1+\frac{3}{5}\lambda_{\theta}}{3 + \lambda_{\theta}}  
      \label{eq2}
\end{equation}

\begin{equation}
       <\cos2\phi> = \frac{\lambda_{\phi}}{3 + \lambda_{\theta}} 
       \label{eq3}
  \end{equation}
  
   \begin{equation}
        < \sin2\theta \cos\phi> = \frac{4}{5} \frac{\lambda_{\theta \phi}}{3 + \lambda_{\theta}} 
        \label{eq4}
   \end{equation}

The polarization parameters $\lambda_{\theta}$, $\lambda_{\phi}$, and $\lambda_{\theta \phi}$ are obtained for 
helicity and Collins-Soper frames by taking the average over  $\cos^{2}\theta$, $\cos2\phi$, and $\sin2\theta 
\cos\phi$. This approach provides an alternative method over the multiparameter fit of Eq.~(\ref{eq1}) to the dimuon angular distribution~\cite{Faccioli:2010kd}.

\subsection{Event Generation in PYTHIA8}
\label{eventgen}
For modeling ultra-relativistic collisions between particles such as electron-electron, electron-positron, proton-proton, and proton-antiproton, one of the commonly used event generators is PYTHIA8. It is quite effective at explaining the LHC results ~\cite{Deb:2018qsl, 
Thakur:2017kpv}. Numerous physical processes are involved in PYTHIA8, including hard and soft scattering, parton distributions, initial and 
final state parton showers, multi-partonic interaction (MPI), string fragmentation, color reconnection, resonance decays, rescattering, and 
beam remnants~\cite{Sjostrand,manual}. In this study, we have used PYTHIA8 to generate $pp$ collisions at $\sqrt{s}$ = 7, 8 and 13 TeV  with 
4C Tune (Tune:pp = 5)~\cite{Corke}. One of the key benefits of PYTHIA8 is the subsequent MPI processes, which, combined with impact 
parameter dependence of collisions, enables the generation of heavy-flavor quarks through 2 $\rightarrow$ 2  hard subprocesses. A detailed 
explanation of all physics processes involved in PYTHIA8 can be found in Ref.~\cite{manual}.\\

 This analysis is performed by generating 1.5 billion events for $pp$ collisions at  $\sqrt{s}$ = 7, 8, and 13 TeV. For our study, we 
 contemplate inelastic and non-diffractive simulated events. So in the total scattering cross section, only the non-diffractive component 
 of all hard QCD processes (HardQCD:all = on) will contribute. Hard processes involve the production of heavy quarks. We have considered 
 color reconnection on (ColourReconnection:mode = on) along with MPI (PartonLevel:MPI = on).  To avoid the divergences of QCD processes in 
 the limit $p_{\rm{T}} \rightarrow 0 $  a transverse momentum cut of 0.5 GeV (PhaseSpace:pTHatMinDiverge = 0.5) is 
 taken. For the production of $J/\psi$ and $\psi$(2S), we use Charmonium:all flag (Charmonium:all = on) in the simulation~\cite{Shao, 
 Caswell, Bodwin} through NRQCD framework. The polarization study of $J/\psi$ and $\psi$(2S) has been performed in the dimuon channel by 
 forcing a $J/\psi$ and $\psi$(2S) to decay into dimuons ($\mu^{+}\mu^{-}$) in the MC simulation. The $J/\psi$ and $\psi$(2S) yields are then 
 obtained through invariant mass reconstruction considering the detector acceptance. This helps in comparing the observations directly with 
the experimental data.\\

To check the compatibility of PYTHIA8 with experimental data, we have used the same tuning as used in our previous works~\cite{Deb:2018qsl, Thakur:2017kpv}, where we have compared the production cross section obtained from PYTHIA8 as a function of transverse momentum and rapidity with the ALICE experimental data of $J/\psi$, and found them to be comparable within uncertainties.


\section{Results and Discussion}
\label{result} 

\begin{figure*}[!]    

   \includegraphics[scale=0.43]{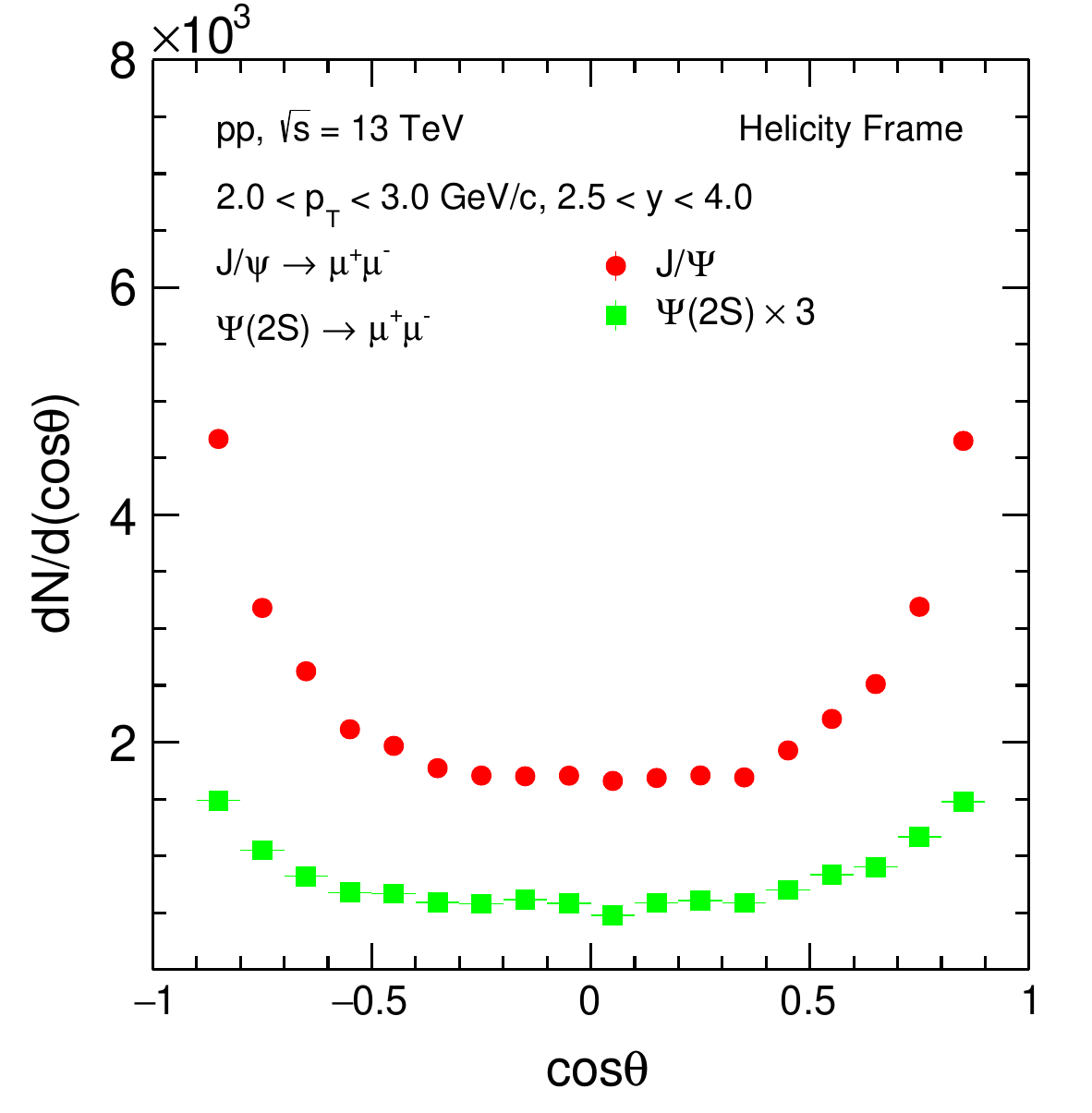}  
   \includegraphics[scale=0.43]{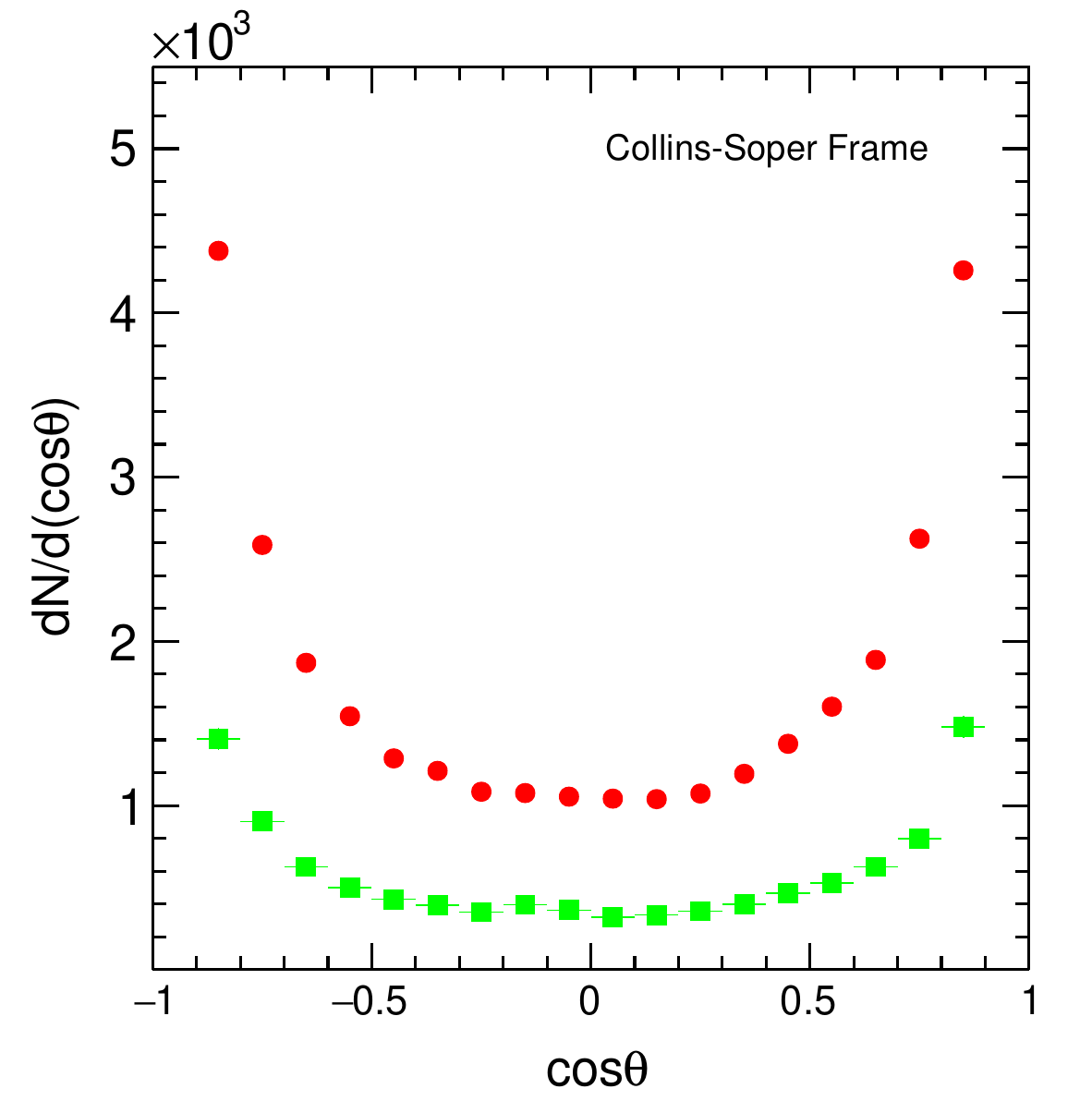}
    \includegraphics[scale=0.43]{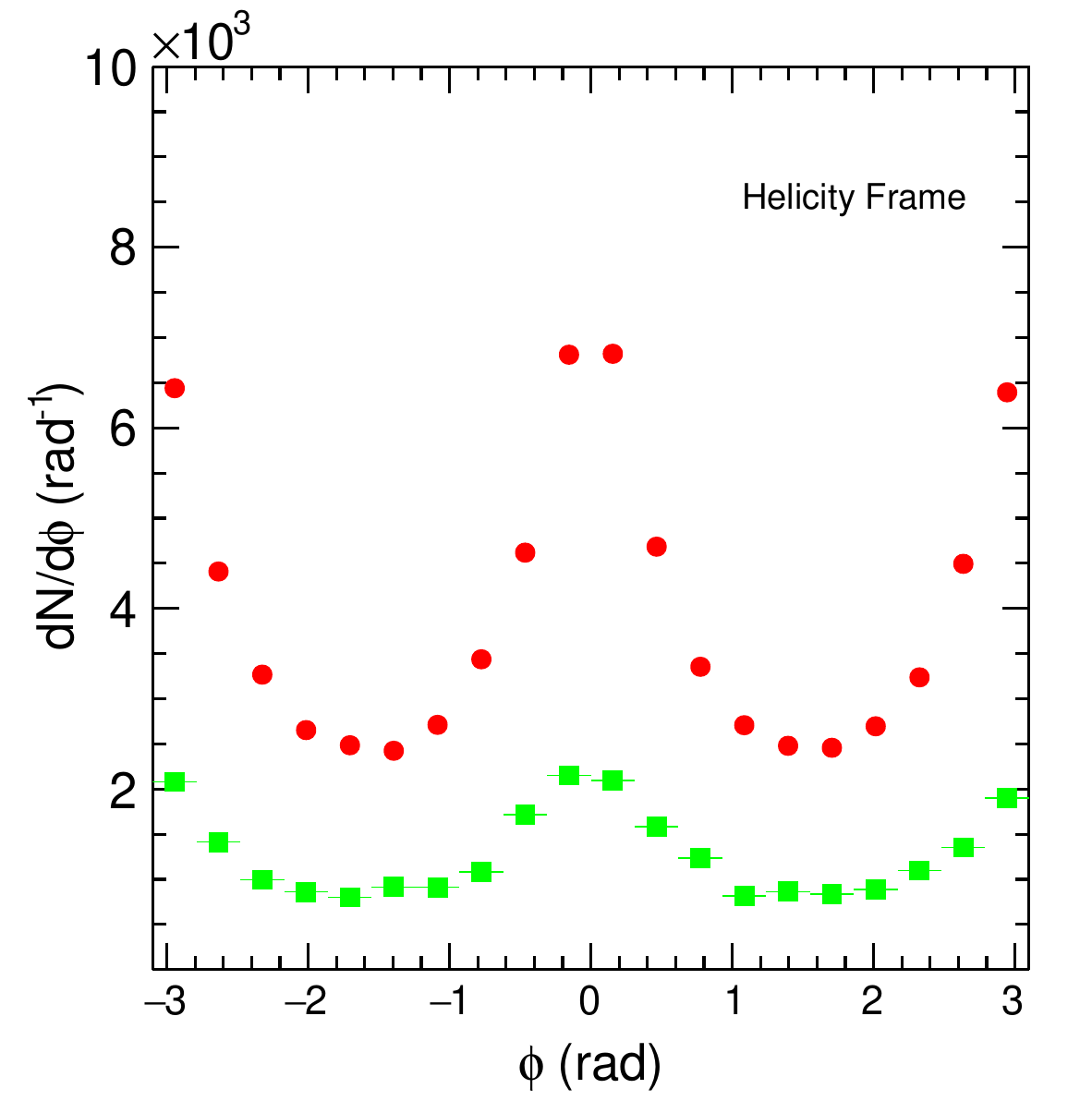}
    \includegraphics[scale=0.43]{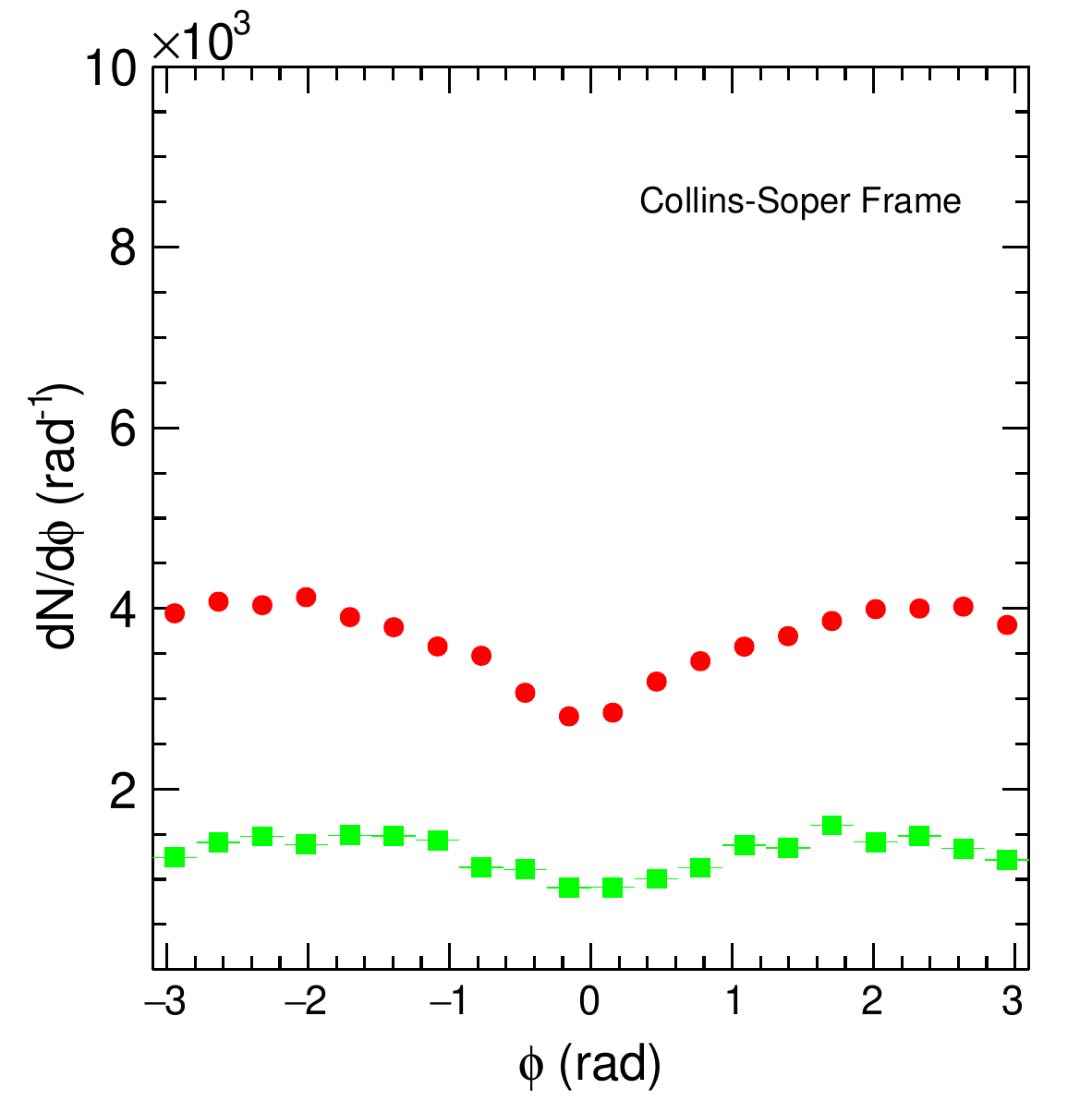}
   
\caption{(Color online) The cosine of the polar angle (upper panel) and the azimuthal angle (lower panel) distribution in $pp$ collisions 
for $J/\psi$ and $\psi$(2S) at $\sqrt{s}$ = 13 TeV in helicity (left panel) and Collins-Soper (right panel) reference frame.}
 \label{fig:1}
\end{figure*}

\begin{figure*}[!]
  \bc    
   \includegraphics[scale=0.8]{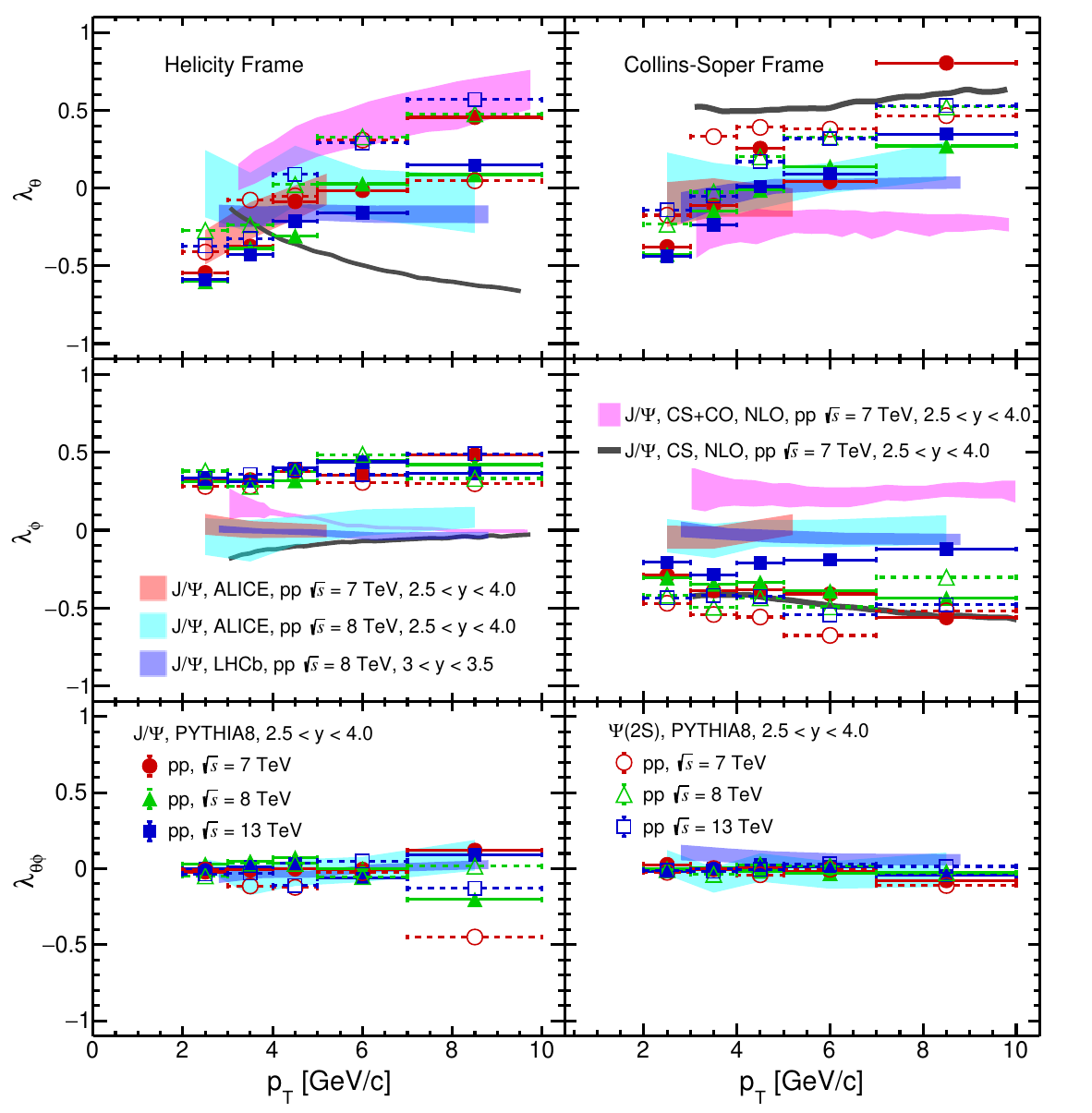}   
   \caption{(Color online) The $J/\psi$ and $\psi$(2S) polarization parameters as a function of transverse momentum for $pp$ collisions at $\sqrt{s}$ = 7, 8, and 13 TeV using PYTHIA8.  The obtained results are compared with the $J/\psi$ polarization measurement in pp collisions from ALICE at $\sqrt{s}$ = 7, 8 TeV, LHCb at $\sqrt{s}$ = 7 TeV and the NLO-NRQCD model predictions in color singlet (CS) and color singlet + color octet (CS+CO) states at $\sqrt{s}$ = 7 TeV in both helicity and Collins-Soper reference frames.}
  \label{fig:2}
\ec
\end{figure*}
\begin{figure*}[!]
  \bc    
   \includegraphics[scale=0.8]{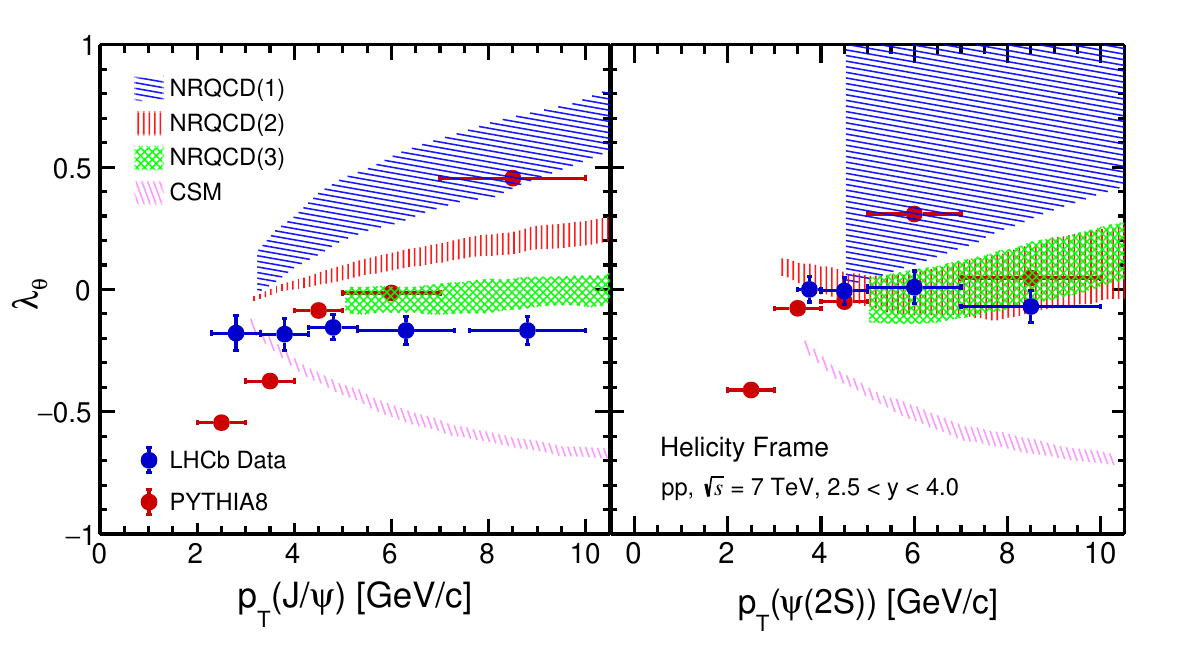}   
   \caption{(Color online) Polarization of $J/\psi$ (left) and $\psi$(2S) (right) as a function of transverse momentum obtained from PYTHIA8 compared with the LHCb experimental data in $pp$ collisions at $\sqrt{s}$ = 7 TeV in the rapidity interval $ 2.5 < y < 4.0$ in helicity frame. The MC simulation results are compared with NLO CSM from~\cite{Butenschoen:2012px} and NLO NRQCD calculation from (1)~\cite{Butenschoen:2012px}, (2)~\cite{Gong:2012ug}, and (3)~\cite{Chao:2012iv, Shao:2012fs}.} 
  \label{fig:3a}
\ec
\end{figure*}

In this section, we discuss the polar and azimuthal angular distribution of dimuons (corresponding to $J/\psi$ and $\psi$(2S)) obtained from the 
PYTHIA8 simulation of $pp$ collisions at $\sqrt{s}$ = 7, 8, 13 TeV. The parameters $\lambda_{\theta}$, $\lambda_{\phi}$ and $\lambda_{\theta \phi}$ are 
obtained from the averaged angular distribution of dimuons using Eq.~(\ref{eq2}), Eq.~(\ref{eq3}), and Eq.~(\ref{eq4}), respectively. The left and 
right columns of Fig.~\ref{fig:1} correspond to the helicity and Collins-Soper frames, respectively. In both the frames, the top panel shows the cosine of 
the polar angle distribution and the bottom panel represents the azimuthal angular distribution for $J/\psi$ and $\psi$(2S) in $pp$ collisions at $\sqrt{s}$ 
= 13 TeV. To obtain these angular distributions, we have considered a $ p_{\rm T}$ bin of $2.0 < p_{\rm T} < 3.0$ GeV/c and a rapidity window
$2.5 < y_{\mu \mu} < 4.0$. Further, the present section is divided into three subsections. The $p_{\rm T}$, charged particle  multiplicity, and rapidity dependence of $\lambda$-polarization parameters (i.e., $\lambda_{\theta}$, $\lambda_{\phi}$, and $\lambda_{\theta \phi}$ ) are discussed in consecutive Secs.~\ref{pt}, ~\ref{mul}, and ~\ref{rap} respectively.
\begin{figure*}[!]
  \bc    
   \includegraphics[scale=0.8]{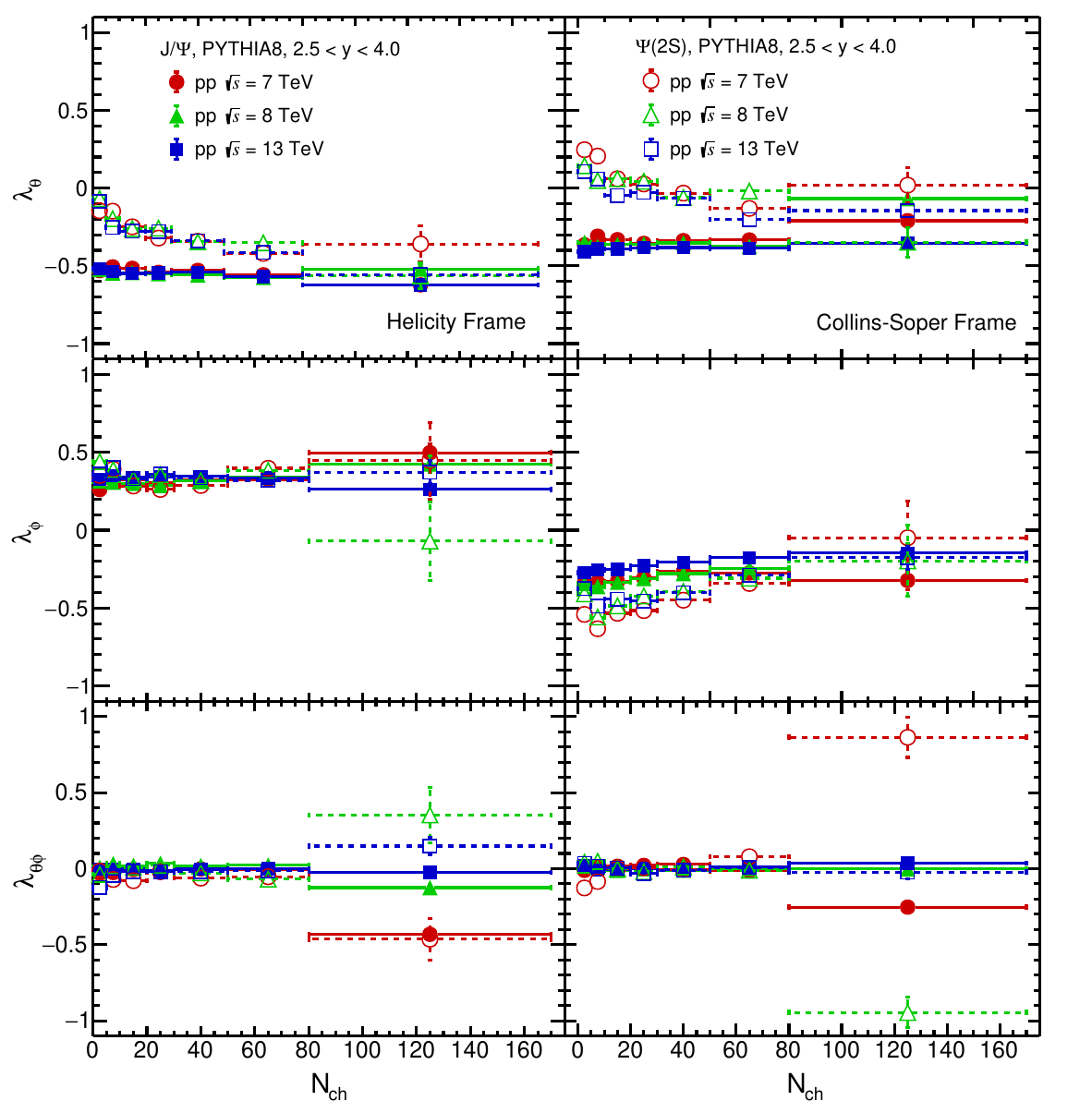}   
   \caption{(Color online) The $J/\psi$ and $\psi$(2S) polarization parameters as a function of charged-particle multiplicity for $pp$ collisions at $\sqrt{s}$ = 7, 8, and 13 TeV using PYTHIA8 in both helicity and Collins-Soper reference frames. } 
  \label{fig:3}
\ec
\end{figure*}

\subsection{Transverse momentum dependence of $\lambda_{\theta}$, $\lambda_{\phi}$, $\lambda_{\theta \phi}$}
\label{pt}

We explore the polarization parameters of $J/\psi$ and $\psi$(2S) as a function of $p_{\rm{T}}$ in $pp$ collisions using PYTHIA8. The 
variation of these parameters at $\sqrt{s}$ = 7, 8, and 13 TeV in both helicity and Collins-Soper reference frame are shown in Fig.~\ref{fig:2}. 
The $J/\psi$ polarization parameters obtained using PYTHIA8  are compared with  the  corresponding experimental data for $\sqrt{s}$ = 7 TeV at LHCb~\cite{LHCb:2013izl} and $\sqrt{s}$ = 7 and 8 TeV at ALICE~\cite{ALICE:2011gej, ALICE:2018crw}.  In addition, the obtained results are compared with the color singlet (CS) and color singlet + color octet (CS+CO) mechanism-based NRQCD, which include the NLO corrections~\cite{Butenschoen:2012px}. The $p_{\rm T}$-interval in PYTHIA8 is chosen in accordance with the ALICE measurement of $J/\psi$ at $\sqrt{s}$ = 8 TeV in both reference frames for all 
energies. The rapidity cut is set to $2.5 < y_{\mu \mu} < 4.0$ in accordance with ALICE detector acceptance. From Fig.~\ref{fig:2}, it is observed that the $\lambda_{\theta}$  parameter indicates a longitudinal polarization at low-$p_{\rm{T}}$ regime, and a transverse polarization at high-$p_{\rm{T}}$, in both the frames of references for $J/\psi$ and $\psi$(2S). This trend qualitatively agrees with the $J/\psi$ polarization measured by ALICE for $\sqrt{s}$ = 7 TeV~\cite{ALICE:2018crw} in the helicity frame. In the obtained result, longitudinal polarization is observed at low-$p_{\rm{T}}$, which decreases towards high-$p_{\rm{T}}$. At lower energy, the HERA-B experiment ~\cite{HERA-B:2009iab} predicts a longitudinal polarization at low-$p_{\rm{T}}$ in the Collins-Soper frame. However, the LO calculation of the NRQCD approach predicts a sizable transverse polarization at high $p_{\rm T}$~\cite{Beneke:1996yw,  Braaten:1999qk, Kniehl:2000nn, Leibovich:1996pa}.\\

Further, it is observed that at low $p_{\rm{T}}$, 
$\psi$(2S) has a comparatively lower longitudinal polarization than $J/\psi$ while the polarization of $\psi$(2S) increases at high $p_{\rm{T}}$. This result 
seems to be apparent because, at low-$p_{\rm{T}}$, the formation of $J/\psi$ through $c-\bar{c}$ is more favourable than $\psi$(2S). As a consequence of this, $\psi$(2S) yield reduces at low $p_{\rm{T}}$, which affects its polarization. This study shows no clear dependence on the center of mass energy on quarkonia polarization in PYTHIA8 for both reference frames. Next,  the $\lambda_{\phi}$ parameter indicates a transverse polarization in the helicity and a longitudinal polarization in the Collins-Soper reference frame. This dissimilarity arises due to differences in the azimuthal angle distribution of dimuons 
around the chosen reference axis in both frames. The finite value of $\lambda_{\theta}$ and $\lambda_{\phi}$ parameters indicates that the probability of finding the $J/\psi$ (or $\psi$(2S)) vector mesons in the three spin states are not equal and hence the emission of their daughter particles is not intrinsically isotropic. We found that the 
PYTHIA8 predicts a relatively higher value of $\lambda_{\phi}$ compared to the experimental data. Although, the $\lambda_{\theta \phi}$ parameter is almost zero for $J/\psi$ and $\psi$(2S) in PYTHIA8, which is consistent with the LHC results as 
displayed in Fig.~\ref{fig:2}.

Fig.~\ref{fig:3a} shows the rapidity-integrated ($2.5 < y < 4.0$) $p_{\rm T}$ dependence polarization measurement of the angular observable $\lambda_{\theta}$ values for $J/\psi$ and $\psi$(2S) in helicity frame, compared with NLO CSM~\cite{Butenschoen:2012px} and different NLO NRQCD calculation~\cite{Butenschoen:2012px,Gong:2012ug,Chao:2012iv,Shao:2012fs}(the NRQCD calculations differ in the experimental data samples used to evalute the long-distance matrix elements). The $p_{\rm T}$ dependence calculations based on the CSM disagree with the experimental data and PYTHIA8 simulation result for $J/\psi$ and $\psi$(2S). For $J/\psi$ resonance, a qualitative agreement between the experimental data and theory is achieved by the NRQCD calculation~\cite{Chao:2012iv, Shao:2012fs}. However, the increasing trend of $\lambda_{\theta}$ parameter with  $p_{\rm T}$ obtained from PYTHIA8 is in agreement up to some extent with NRQCD(1)~\cite{Butenschoen:2012px} and NRQCD(2)~\cite{Gong:2012ug} calculations. Furthermore, for $\psi$(2S), the LHCb data and PYTHIA8 favours the calculation from Ref.~\cite{Chao:2012iv,Shao:2012fs} and from Ref.\cite{Gong:2012ug} at low $p_{\rm T}$. The main difference between the polarization measurement of $J/\psi$ and $\psi$(2S) in PYTHIA8 could be due to their different masses. In experimental data, the $J/\psi$ polarization can come from directly produced $J/\psi$ mesons and from those produced in the decay of heavier charmonium states. However, the polarization of $\psi$(2S) will remain unaffected by feed-down decays of heavier charmonia.\\

\begin{figure*}[!]
  \bc      
    \includegraphics[scale=0.68]{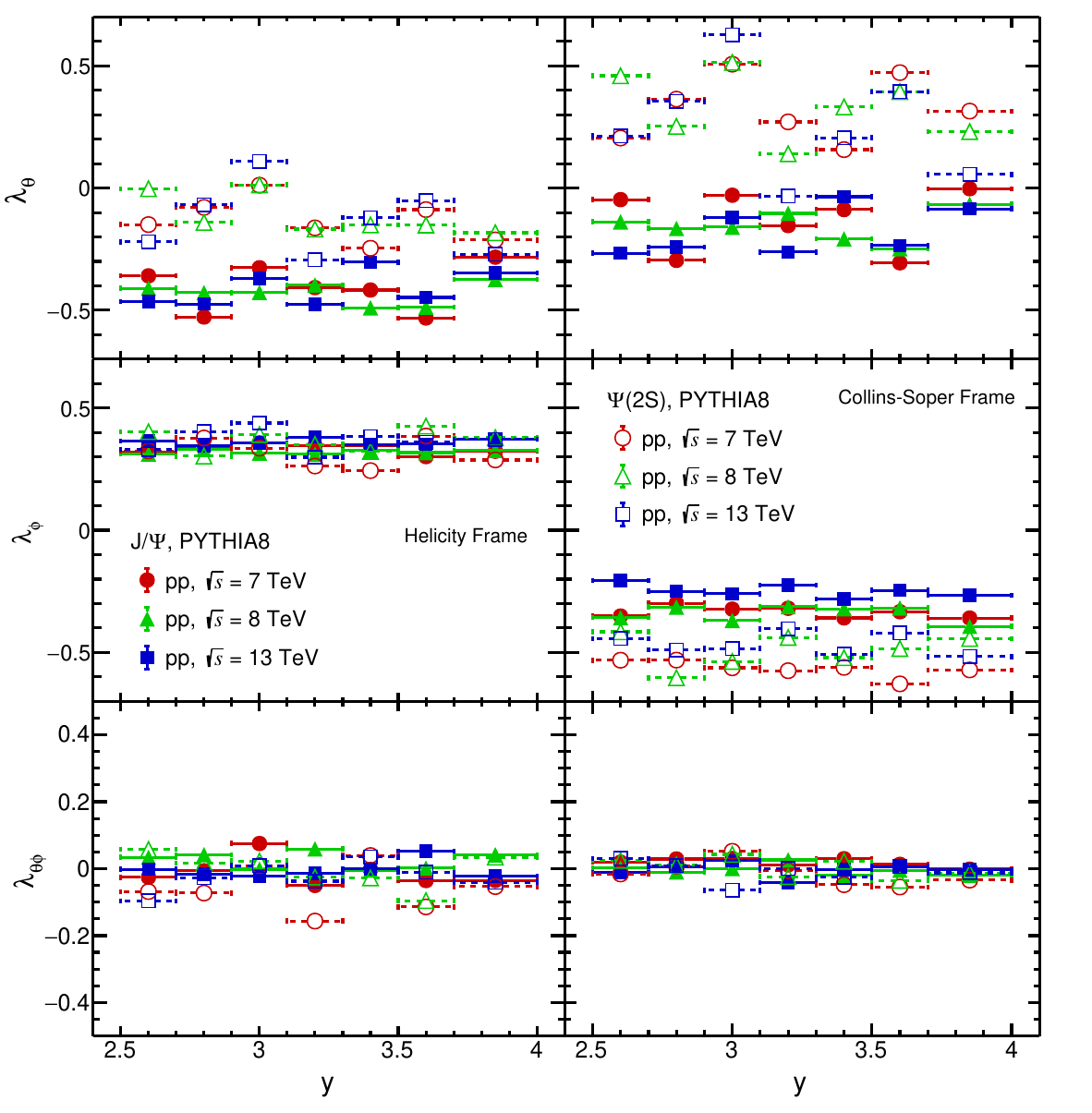}
   \caption{(Color online) The $J/\psi$ and $\psi$(2S) polarization parameters as a function of rapidity for $pp$ collisions at $\sqrt{s}$ = 7, 8, and 13 TeV using PYTHIA8 in both helicity and Collins-Soper reference frames.}
  \label{fig:4}
\ec
\end{figure*}
\subsection{Charged-particle multiplicity dependence of $\lambda_{\theta}$, $\lambda_{\phi}$, $\lambda_{\theta \phi}$}
\label{mul}

The charged particle multiplicity-dependent study of charmonia polarization may reveal the underlying dynamics associated with the particle density of the system produced in $pp$ collisions. 
Figure~\ref{fig:3} shows the charged particle multiplicity dependence of polarization parameters for $J/\psi$ and $\psi$(2S) mesons in $pp$ collisions at 
$\sqrt{s}$ = 7, 8, and 13 TeV. The charged-particle multiplicity classes used in the present analysis are taken from Ref.~\cite{Deb:2018qsl}. The experimental 
study in this regard is reported by CMS collaboration for three $\Upsilon(nS)$ states in $pp$ collisions at  $\sqrt{s}$ = 7 TeV~\cite{CMS:2016xpm}. From the theoretical front, the relative multiplicity dependence $\left(\frac{dN_{ch}}{d\eta}/\left<\frac{dN_{ch}}{d\eta}\right>\right)$ study of polarization parameters 
for $J/\psi$ in helicity and Collins-Soper frame is studied in CGC+NRQCD approach~\cite{Stebel:2021bbn} in $pp$ and $p$-Pb collisions at $\sqrt{s}$ = 13 and 
8.16 TeV, respectively. From Fig.~\ref{fig:3}, we observe that the polarization parameter $\lambda_{\theta}$ indicates that the degree of longitudinal 
polarization increases towards high multiplicity for $\psi$(2S) meson, while for $J/\psi$ the longitudinal polarization remains almost constant from low-
to high multiplicity in both reference frames. The $\lambda_{\phi}$ shows a transverse polarization in the helicity frame and a longitudinal 
polarization in the Collins-Soper reference frame for $J/\psi$ and $\psi$(2S) for all multiplicity classes. The polarization parameter, $\lambda_{\theta 
\phi}$, is negligible for $J/\psi$ and $\psi$(2S) at low multiplicities, while at higher multiplicities, it has non-zero values. The present study indicates that the 
charmonia polarization weakly depends on the center of mass collision energy ($\sqrt{s}$).

\subsection{Rapidity dependence of $\lambda_{\theta}$, $\lambda_{\phi}$, $\lambda_{\theta \phi}$}
\label{rap}

 In this section, we explore the rapidity dependence study of $\lambda$-parameters for $J/\psi$ and $\psi$(2S) mesons in $pp$ collisions at $\sqrt{s}$ = 7,  8, and 13 TeV, as shown in Fig.~\ref{fig:4}. The  $J/\psi$ polarization for various rapidity bins is reported in the LHCb experiment for $pp$ collisions at 
 $\sqrt{s}$ = 7 TeV~\cite{LHCb:2013izl} and data show a small polarization within the uncertainties. In Fig.~\ref{fig:4}, the $p_{\rm{T}}$-integrated 
 polarization is obtained for the rapidity range 2.5 to 4.0, with a step of 0.3. The top panel of Fig.~\ref{fig:4} shows that the 
 degree of longitudinal polarization for $J/\psi$ is larger than $\psi$(2S) in the helicity frame. It observed that in the Collins-Soper frame, $\psi$(2S) is 
 transversely polarized while $J/\psi$ is longitudinally polarized. However, as observed in the present study, there is no clear dependence of 
 $\lambda_{\theta}$ parameter with rapidity in both reference frames. Similar to $p_{\rm{T}}$ and charged-particle multiplicity dependence, the $\lambda_{\phi}$ 
 as a function of rapidity shows a positive value for polarization in the helicity frame and a negative value of polarization in the Collins-Soper frame for 
 $J/\psi$ and  $\psi$(2S). The $\lambda_{\theta \phi}$ parameter shows almost zero polarisation with rapidity irrespective of the particles under consideration, 
 the chosen center of mass energy, and the reference frame.\\

The finite polarization of $J/\psi$ and $\psi$(2S) in PYTHIA8 may have several possible sources, a few of which are discussed below. In HICs (Au+Au, Pb+Pb, etc.), the potential sources of hadrons polarization are the vorticity fields, electromagnetic fields, and a strong vector meson force field~\cite{STAR:2017ckg, Han:2017hdi, Gao:2012ix, Sheng:2020ghv, Sheng:2022wsy}. However, the formation of vorticity due to the initial orbital angular momentum and the electromagnetic fields is almost negligible in pp collisions. Therefore, the charmonium polarization in ultra-relativistic pp collisions is an expected consequence of the involved production mechanisms~\cite{Faccioli:2022peq, Faccioli:2010kd}. As PYTHIA8 considers LO NRQCD mechanisms for quarkonia production, it can be anticipated that the finite polarization of charmonium states is due to the QCD processes involved in PYTHIA8. In this work, we use the flag HardQCD:all = on, which favors the production of heavy flavors through all hard-QCD $2 \rightarrow 2$ scattering processes. It involves pair creation [q$\bar{q}$ (gg) $\rightarrow$ c$\bar{c}$], flavour excitation, and gluon splitting processes~\cite{manual}. Moreover, a few basic principles, such as: i) helicity conservation in the dilepton decays; ii) rotational covariance of angular momentum eigenstates; and iii) conservation of parity and its violation, govern the polarization of charmonium states~\cite{Faccioli:2022peq, Faccioli:2010kd}. In addition, we use the flag Charmonium:all = on, which includes the production of charmonium and its higher excited states via the color-singlet and color-octet mechanism of NRQCD, requiring information on long-distance NRQCD matrix elements for the various wavefunctions involved in PYTHIA8~\cite{manual}. Further, the role of MPI on the decay angular distribution of dimuon is investigated in Appendix~\ref{appe1}. \\

Another source of polarization is the inhomogeneous expansion of the medium, resulting in an anisotropic flow in the transverse plane. The second anisotropic flow-coefficient (i.e. elliptic flow) induced polarization is observed for $\Lambda$-hyperons in HICs~\cite{Becattini:2017gcx, STAR:2019erd}. In $pp$ collisions, a similar flow-like phenomenon is obtained in PYTHIA8 due to the considered multi-parton interactions (MPI) and color reconnection mechanism~\cite{OrtizVelasquez:2013ofg}. In the present work, we use the same tune MPI = on and CR = on; as an outcome, the flow-like effect indicates another possible source of $J/\psi$ and $\psi(2S)$ polarization.\\

It is observed that PYTHIA8 serves as an effective model for jet studies in $pp$ collisions with MPI and CR mode on~\cite{ALICE:2023oww}. The jet-like fluctuation in PYTHIA8 is also a possible source of polarization for hadrons. Because, if a jet is produced in the medium it deposits energy and creates a smoke-loop-type vortex in the jet plane. As a consequence, it induces polarization to the particles associated with the jets~\cite{Betz:2007kg}. Likewise,  charmonia can also get polarized due to such phenomena involved in the production processes.\\

Along with the discussed polarization mechanisms, there are other possible sources of polarization for charmonium states, but it is still a matter of investigation whether the spin orientation or polarization occurs at the partonic level and then is transferred to the bound state or directly to the bound state. Based on the present study, we may infer that NRCQD-based production mechanism give rise to the possible source for the partonic level polarization and then is transferred to the bound state. However, the flow-like effect is attributed to the direct polarization of the bound state.

\section{Summary}
\label{summary}
In this work, we have studied the polarization parameters for $J/\psi$ and $\psi$(2S) from the angular distribution of dimuons in $pp$ collisions at LHC 
energies using PYTHIA8. The important observations of this paper are summarized below:

\begin{enumerate}
    \item The $\lambda_{\theta}$, $\lambda_{\phi}$, $\lambda_{\theta \phi}$ are obtained in the helicity and Collins-Soper reference frames in the rapidity interval $2.5 < y_{\mu \mu} < 4.0$.

    \item It is observed from the  $\lambda_{\theta}$ parameter that $J/\psi$ and $\psi$(2S) are longitudinally polarized at low-$p_{\rm{T}}$ and transversely polarized at high-$p_{\rm{T}}$ in both the reference frames.

    \item  The  $\lambda_{\phi}$ parameter indicates the longitudinal polarization in helicity and transverse polarization in the Collins-Soper frame for $J/\psi$ and $\psi$(2S) across all energies. The $\lambda_{\theta \phi}$ parameter values are close to zero as a function of $p_{\rm{T}}$. 

    \item The multiplicity dependence study of $\lambda_{\theta}$ parameter shows the degree of longitudinal polarization increases with charged particle multiplicity for $\psi$(2S), while the behavior of longitudinal polarization of $J/\psi$ stays constant with charged particle multiplicity, which needs attention from an experimental point of view.

    \item  In this study, we observe no clear dependence of $\lambda_{\theta}$ parameter with rapidity. However the  $\lambda_{\phi}$, $\lambda_{\theta}$ show almost constant polarization with rapidity both for $J/\psi$ and $\psi$(2S) in helicity and Collins-Soper reference frames. In the future, ALICE 3 set up with a wider kinematics acceptance range of muon spectrometer, the rapidity dependence study of polarization parameters would be an interesting topic.

    \item It is essential to mention that the polarization results obtained in this analysis as a function $p_{\rm{T}}$, $N_{ch}$, and $y_{\mu \mu}$ consider only the production of $J/\psi$ and $\psi$(2S), without taking the feed-down from higher excited states. The investigation of charmonia polarization by taking the feed-down from higher resonances is a future scope. In other words, the newly upgraded Muon Forward Tracker (MFT) detector in the Run 3 and Run 4 setup of ALICE 2 to the muon Spectrometer, having better vertexing and precise measurement capabilities, helps us to separate for prompt and non-prompt charmonium states. Therefore, the investigation of polarization parameters for prompt and non-prompt charmonium states in both the experiment and MC simulation would be an intriguing subject to test.

     \item Since we used PYTHIA8, which incorporates pQCD and NRQCD-based processes, our results overestimate $J/\psi$ polarization in some $p_{\rm{T}}$ bins as compared with experimental data. It suggests that there might be an interplay between these fundamental processes in the realistic scenario. On the other hand, there might be some other processes responsible for net charmonia polarization. Therefore, charmonia polarization in ultra-relativistic $pp$ collisions requires a thorough study using theoretical models confronted with the experimental results.    
     
\end{enumerate}

\section*{Acknowledgement}

B. S. acknowledges the financial aid from CSIR, Government of India. SD acknowledges the financial support under the Post-Doctoral Fellowship of CNRS at IJCLAB, Orsay (France). The authors gratefully acknowledge the DAE-DST, Government of India funding under the mega-science project "Indian Participation in
the ALICE experiment at CERN" bearing Project No.
SR/MF/PS-02/2021-IITI (E-37123).
\vspace{10.005em}

\appendix

\section{ROLE OF MPI ON POLARIZATION PARAMETERS}
\label{appe1}
\begin{figure*}[t]
  \bc    
   \includegraphics[scale=0.8]{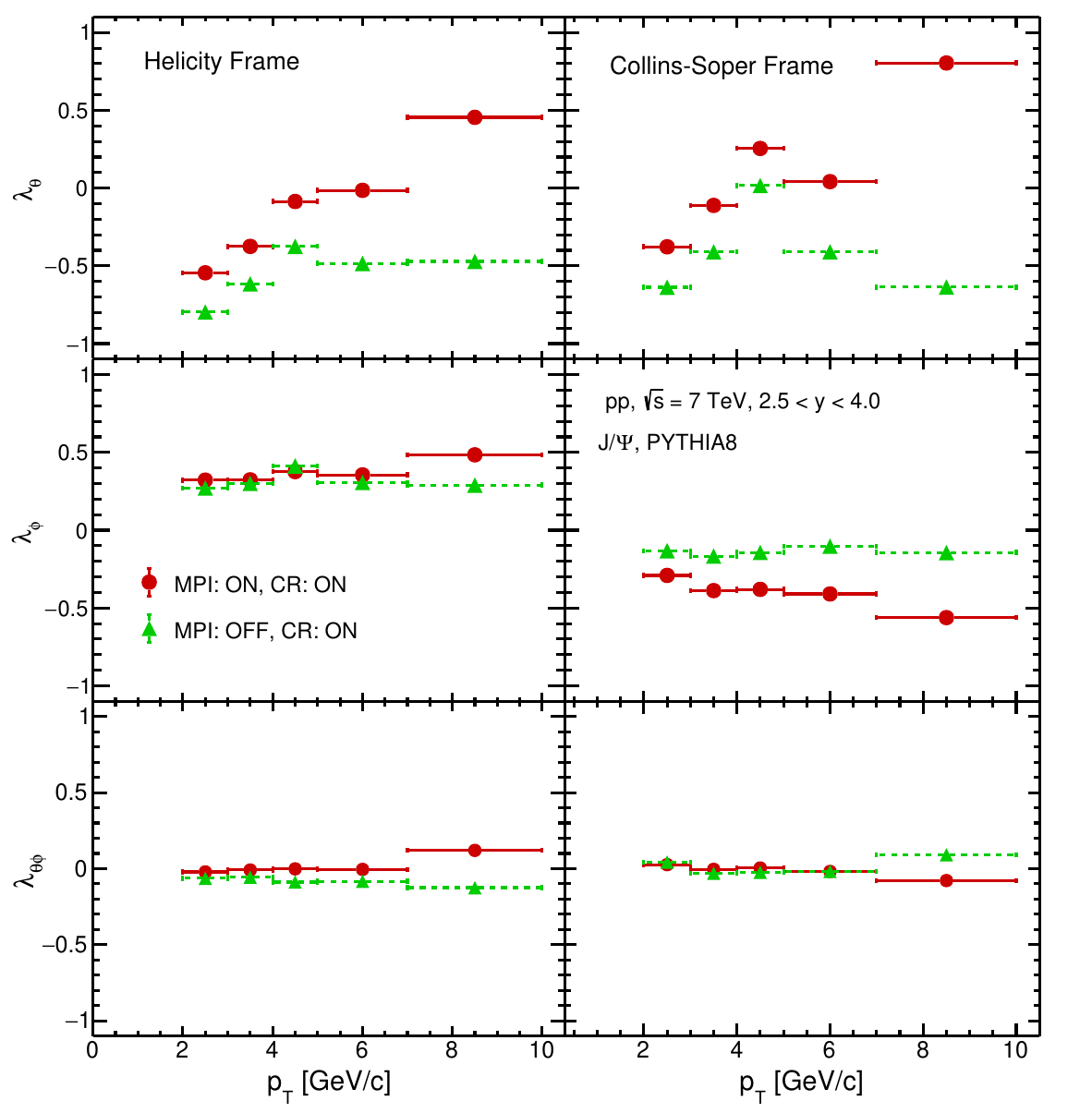}   
   \caption{(Color online) The $J/\psi$ polarization parameters as a function of transverse momentum with \enquote{MPI: ON, CR: ON} (Red Marker) and \enquote{MPI: OFF, CR: ON} (Green Marker) for $pp$ collisions at $\sqrt{s}$ = 7 using PYTHIA8 in helicity and Collins-Soper reference frames.}
  \label{mpi}
\ec
\end{figure*}

In this study, our investigation centered on charmonia state polarization within proton-proton collisions simulated through PYTHIA8. Multiple parton interactions (MPI) in PYTHIA8 primarily affect parton distribution and activity. However, their indirect influence potentially leads to alterations in the angular momentum distribution among final-state particles due to modified partonic interactions and subsequent hadronization processes. These changes can manifest as variations in polarization parameters, particularly $\lambda_{\theta}$, which are sensitive to the angular momentum states of generated particles. Fig.\ref{mpi} highlights our investigation of the impact of MPI (with on/off settings) on polarization parameters. Notably, our focus rests on the observed behavior of $\lambda_{\theta}$, revealing distinctive trends: when MPI is enabled (on), $\lambda_{\theta}$ displays an initial negative trend, transitioning to positive values with increasing transverse momentum. Conversely, in the absence of MPI, $\lambda_{\theta}$ consistently maintains negativity with increasing transverse momentum. These trends persist consistently in both the helicity frame and the Collins-Soper frame. The intriguing shift from negative to positive $\lambda_{\theta}$ values with MPI compared to the sustained negativity without MPI suggests a significant influence of multiple parton interactions on polarization phenomena. Possible explanations for these observations might involve variations in partonic interactions and their consequential indirect effects on angular momentum dynamics. This underscores the imperative need for further theoretical and experimental investigations to unravel the underlying mechanisms driving polarization in high-energy collisions.

\end{document}